Discussion Paper:Theory of opinion distribution in human relations where trust and distrust mixed(2020)

# Discussion of the Effect of Inter-group Sub-groups Using a Consensus Model Incorporating External Effective or Immobile Magnetic Fields

Yasuko Kawahata [†]

Faculty of Sociology, Department of Media Sociology, Rikkyo University, 3-34-1 Nishi-Ikebukuro,Toshima-ku, Tokyo, 171-8501, JAPAN.
`ykawahata@rikkyo.ac.jp,kawahata.lab3@damp.tottori-u.ac.jp`

**Abstract:** Individuals belong to certain social groups in search of a sense of belonging, pride, stability, and significance. Perceiving the group to which one belongs as an "in-group" and other groups as "out-groups" often leads to harmful and discriminatory attitudes. In-group consciousness reinforces a sense of unity within the group and promotes commitment to group goals and problem solving. Identification with the in-group also shapes the social cognitive framework (norms, values, and beliefs) that determine group behavior. In fact, identification with an in-group often leads to prejudice, ethnocentrism, stereotyping, and discrimination, even in the absence of physical conflict or hostility. Social scientists have conducted thousands of empirical studies to elucidate the mechanisms behind these prejudices and discriminations and the social conflicts they generate. These studies are essential to understanding the processes by which group membership and self-categorization create prejudice and discrimination, which in turn lead to social conflict. However, there remain many unanswered questions about how in-groups and out-groups can move beyond conflict to build harmony and avoid social conflict. According to existing research, it is difficult to establish harmonious relationships between in-groups and out-groups. This study proposes an approach using opinion dynamics theory and social simulation to examine these issues. We examine the possibility of simulating the movement of opinions between and within groups and applying the considerations to cases of social conflict. The model analyzes the severity of conflict within a society with two groups on the basis of intragroup and intergroup trust. The simulation in this paper is based on the theory of opinion dynamics, which incorporates both group opinion A and the opposite case of group opinion B in human relationships. It was confirmed that the aspect of consensus formation depends on the ratio of the trust and distrust coefficients. In the present study, the ratio of trust to distrust tended to change like a phase transition around 55%, and we wondered whether a similar phenomenon could be confirmed for large-scale cases. In previous case studies, this trend was observed from $N = 300$ to $N = 10000$. By examining the extent of this phenomenon on a social scale, we hope to examine simulation items for consensus building, such as consideration of the sensitivities of topics in online and offline opinion formation. In particular, this paper will focus on the discussion focus on consensus building between two groups.

## 1. Introduction

Throughout world history, social conflicts have occurred frequently, and their causes and solutions are complex. Case analyses of social conflicts can help us find clues to prevent future crises. This paper focuses on the dynamics of conflict between two groups and explores its mechanisms using simulations. According to social identity theory, prejudice and discrimination are the primary causes of social conflict, not individual relationships or psychological processes as direct causes.

Social identity theory, proposed by Tajfel and Turner et al. (1979), suggests that individuals may hold negative stereotypes and prejudices toward out-groups in order to maintain a positive self-image toward the in-group to which they belong. The French Revolution, which occurred between 1789 and 1799, was an outpouring of dissatisfaction with absolute monarchy and the social class system, and prompted the rise of modern democracy and the establishment of the rule of law. Tocqueville (1856) has described his views on this. On the other hand, McPherson (1988) notes that the Civil War of 1861-1865, caused by the conflict over the existence of slavery in the United States, resulted in emancipation and national unity, but racial tensions left deep scars.

With regard to the Arab Spring, Howard and Hussain (2013) and others note the important role of social media from 2010 to 2012, when citizen protests against political crippling and economic deprivation led to the overthrow of successive governments. In addition, the Rwandan civil war and genocide led to heightened ethnic tensions between 1990 and 1994, resulting in a catastrophic massacre of approximately 800,000 Tutsis and moderate Hutus.

The need for prompt intervention by the international community was addressed in Gourevitch (1998) on this incident. The breakup of the former Yugoslavia and the subsequent conflict occurred between 1991 and 2001, an event that illustrates the complexity of ethnic self-determination, territorial sovereignty, and international intervention, as reported in Glenny (1996). The situation in Russia and Ukraine also suggests that the annexation of Crimea in 2014 and the ongoing conflict in Eastern Ukraine have created new geopolitical tensions, threatening the principles of international law and the preservation of regional stability, to Sakwa (2015). The Iraq War (2003-2011) and the Syrian Civil War (since 2011) highlight the difficulties of intervention in international relations and its long-term effects of social and economic destruction. These are mentioned in Friedman (2009) and Phillips (2016), respectively. The democratization movement in Hong Kong, characterized by the tension between the growing demand for democracy under China's one country, two systems and the central government's increasing political influence, is highlighted in Chan, J. and Chow, E. (2020), especially the series of protest movements beginning with the 2014 Umbrella Revolution. Furthermore, we highlight in Ortmann, S. (2015) the importance of large-scale protests in 2019. With regard to East Timor's independence movement, Budiardjo, C. and Liong, L. W. (2001) state that after more than 25 years of Indonesian occupation, a United Nations-led referendum in 1999 paved the way for independence, and Braithwaite, J., Charlesworth, H . and Soares, A. (2012), citing this case as a rare example of successful international intervention. As for other cases in Asia, the Cambodian conflict, as described by Chandler, D. P. (1991) and others, was a protracted war characterized by political strife, intervention by foreign powers, and massacres by the Khmer Rouge. And the Vietnam War was a Cold War conflict, which Young, M. B. (1991) et al. point out was characterized by its civilian impact and foreign intervention. Tensions in China's Uyghur Autonomous Region, where Uyghurs oppose religious and cultural oppression and unfair economic treatment, are noted in Becquelin, N. (2004) in the international community's debate on China's minority policies.

After more than 25 years of Indonesian occupation and repression, East Timor's independence movement began its path to independence with a United Nations-led referendum in 1999. This is a rare example of successful intervention by the international community and illustrates the importance of international efforts to resolve regional conflicts. The issue of East Timor's independence has a history that includes the invasion and subsequent occupation by Indonesia and the independence movement that resulted in an internationally supported independence vote in 1999. This event is an example of the role that international intervention can play in resolving conflicts in the region. In other Asian cases, the Cambodian conflict was a protracted and complex war characterized by political strife, intervention by foreign powers, and ultimately the Khmer Rouge genocide under Pol Pot. The conflict marked a serious division within the country and later led to international intervention. The Vietnam War occurred within the context of the Cold War, a conflict between North Vietnam (later the Socialist Republic of Vietnam) and South Vietnam (the Republic). The war claimed many civilian lives, caused deep domestic divisions, and was characterized by foreign intervention, including that of the United States. In the Uyghur Autonomous Region of China, there have been long-standing tensions between the Uyghurs and the Chinese government. Uyghurs oppose religious and cultural oppression, unfair economic treatment, and lack of political freedom. The issue has attracted the attention of international human rights organizations and has sparked a broad debate on China's ethnic minority policies.

## 1.1 The phenomenon of intra-group identification causing prejudice and discrimination: the field of social psychology

Tajfel and Turner (1979) proposed the social identity theory. This theory suggested that individuals hold negative stereotypes and prejudices toward out-groups in order to maintain a positive self-image toward in-groups; Tajfel and Turner et al. were the first to systematically introduce the basic elements of social identity theory. Exploring the roots of identity-based intergroup conflict, they argue that individuals may harbor prejudice or hostility toward other groups in order to enhance their self-esteem, and their 1986 study details how social identity theory is actually applied to intergroup behavior. It discusses the dynamics between in-groups and out-groups and how individuals perceive and evaluate their groups. In addition, Allport's (1954) contact hypothesis indicates that cooperative and equal status contact between in-group and out-group members may reduce prejudice, but suggested that the reverse is also true.Pettigrew (1998) has conducted numerous studies on prejudice and discrimination and conducted a number of experiments examining how in-group identification shapes and reinforces negative attitudes toward out-groups, which attracted much interest. Brewer (1999) explored the principle of in-group bias and theorized how group identity can lead to prejudice toward out-groups.

## 1.2 The severity of the conflict within a society where two groups are present

The Rwandan Genocide (Prunier, G. [1995]. The Rwanda Crisis: History of a Genocide.) and the Bosnian conflict (Glenny, M. [1996]. The Fall of Yugoslavia: The Third Balkan War.) are perhaps the most tragic examples of intergroup conflict.

Donald L. Horowitz (1985) analyzes patterns and causes of conflict between ethnic groups. He identified cultural differences, economic competition, and political inequality as the main factors that exacerbate conflict.Ted Robert Gurr (1993) analyzed how political discontent and intergroup inequality evolve into violence and proposed the theory of relative deprivation. This theory posited that the greater the gap between expectations and reality, the greater the likelihood that a group will revolt.

## 1.3 Case Study of Political Unrest

Robert D. Putnam (2006) analyzed the impact of diversity on social capital and community cohesion. His study showed that in the short term diversity can reduce social solidity and trust, which can be a factor in generating political unrest. Paul Collier and Anke Hoeffler (2004) studied the economic and social conditions that cause civil wars and explained how economic factors and organizational inequalities can cause political unrest and violent conflict.

I. William Zartman (1995) edited a collection of essays on state collapse and the political unrest that accompanies it, exploring how the loss of government legitimacy can cause serious social conflict.

James D. Fearon, David D. Laitin, et al. (2003) analyzed how ethnic identity influences the occurrence of civil war and suggested intra-societal conflict as a factor in political instability.

## 1.4 Cases of Economic Losses

Amartya Sen (1999) rethought development in terms of freedom and analyzed social losses and conflicts caused by lack of economic freedom, such as poverty, hunger, and lack of education.

Joseph E. Stiglitz (2003) explored the mechanisms by which globalization creates economic inequality and social unrest, especially in developing countries, and discussed how this leads to conflict and economic loss.

Paul Collier (2007) analyzed the economic challenges faced by the world's poor and the resulting conflicts and showed how economic stagnation leads to political instability and conflict.

Branko Milanovic (2016) proposed a new measure of economic inequality and examined how global inequality contributes to economic losses, including conflict.

## 1.5 Case Study of Deteriorating International Relations

Graham Allison (1971) used the Cuban Missile Crisis as a case study to analyze how state decision-making processes can affect international relations and increase tensions, and Kenneth Walz (1979) analyzed how the structure of the international system determines state behavior and how it can contribute to the deterioration of international relations explained how this may be the case. Samuel Huntington (1996) argued and determined that clashes between different civilizations will be one of the main causes of the deterioration of international relations in the 21st century.

John Mearsheimer (2001) discussed how strategic competition among great powers can worsen international relations and analyzed it from the perspective of realist theory. Joseph Nye (2001) discussed the use of non-power influence to prevent international relations from deteriorating through the concept of soft power.

## 1.6 Opinion on the Internet

The division of opinion on the Internet is identified through this event as an emerging form of social conflict. In particular, the Black Lives Matter movement (2013-) is an example of how protests against police violence against black people showed social solidarity and unity of action, and the movement continues to seek to reform law enforcement and raise social awareness (Anderson, 2016). This movement is also a prime example of how social media can highlight social issues and enable collective action across divisions of opinion. By examining these examples of social conflict, we can understand the causes from multiple perspectives and explore avenues for finding effective solutions. Through these conflicts, we can also gain deeper insights into social inequalities and injustices, which will hopefully lead to the development of policies and theories to prevent future conflicts and create a more equitable and peaceful society.

Individuals belong to specific social groups for a sense of belonging, pride, stability, and meaning. Perceiving the group to which one belongs as an "in-group" and other groups as "out-groups" often leads to harmful and discriminatory attitudes. In-group consciousness reinforces a sense of unity within the group and promotes commitment to group goals and problem solving. Identification with the in-group also shapes the social cognitive framework (norms, values, and beliefs) that determine group behavior. Indeed, identification with an in-group often leads to prejudice, ethnocentrism, stereotyping, and discrimination, even in the absence of physical conflict or hostility. For example, the split between the Republican and Democratic parties in the United States, or the conflict between supporters of former President Trump and other citizens, these can be understood through social identity theory.

## 1.7 Study of In-group and Out-group Dynamics

This study by Bloomer (1958) has provided a theoretical framework for how racial prejudice arises from group status consciousness.

This classic work on ethnocentrism, or ethnocentrism, by LeVine, R. A. (1972) and others, has focused on theories and attitudes related to in-group and out-group conflict.

Gordon Allport (1954) used a classical research approach to investigate the underlying causes of prejudice and how it operates within society. The work of Muzaffer Sheriff (1966) experimentally explored the psychology of intergroup conflict and cooperation and practiced challenging research, including the famous Robbers Cave Cave experiment. Sherriff's experiments empirically demonstrated the effects of competition and cooperation on intergroup relations and made important contributions to the theory of intergroup conflict. This work by Jim Sidanius and Felicia Pratt (1999) introduces social control theory and explains how certain groups maintain social hierarchies and oppress out-groups. Marilyn Brewer (1999), on the other hand, offered the perspective that prejudice does not necessarily stem from hatred of the out-group, but can also stem from affection for the in-group. She proposed the idea that prejudice does not arise from hatred of the out-group, but is linked to a strong attachment to the in-group. Sidanius, J (1999) et al. (1999) explained how social hierarchy and oppression shape intergroup relations, which they called "social dominance theory" (Sidanius, J, et al., 1999). Haslam (2006) explained how denial of humanity justifies prejudice and discrimination against out-groups.

## 1.8 Research on Conflict Resolution

By Lederach, J. P. (1997) et al. Research on conflict resolution in conflict involves initiating dialogue between the parties to the conflict and facilitating the process of reconciliation. It has been suggested that this often requires the mediation of a third party.

Thomas F. Pettigrew (1998) developed the contact hypothesis that contact between different groups increases mutual understanding and decreases prejudice. He proposed that in societies where intergroup conflict is severe, cooperative interactions between different groups can be effective in mitigating conflict.

Social scientists have conducted thousands of empirical studies to elucidate the mechanisms behind these prejudices and discriminations and the social conflicts they create. These studies are essential to understanding the processes by which group membership and self-categorization create prejudice and discrimination, which in turn lead to social conflict. However, there remain many unanswered questions about how in-groups and out-groups can move beyond conflict to build harmony and avoid social conflict.

According to existing research, it is difficult to establish harmonious relationships between in-groups and out-groups. This study responds to these challenges by proposing a new approach using opinion dynamics theory and social simulation. We simulate the movement of opinions between and within groups and apply it to a case study of social conflict. The model analyzes the severity of conflict within a society with two groups on the basis of intragroup and intergroup credibility.

## 1.9 Inference and Hypothesis of Social Phenomena

Subgroups affected by items in invariant magnetic field $F$ Influence on $A$ and $B$

### 1. factors that can contribute to the inference of social phenomena

Exchange of Opinions and Interaction: The model emphasizes the interaction between individuals and the exchange of opinions as a key factor in opinion dynamics. Specifically, the trust matrix $d$ indicates the strength and direction of this interaction, which we hope will help us understand patterns of opinion spread and change on social networks.External influences: External influences, represented as $F$, mimic the influence of information, opinions, media, etc. from outside the social network. By doing so, we hope to understand how external information sources influence opinion formation.

Initial distribution of opinions: The initial distribution of opinions can have a significant impact on the simulation results. By investigating how different initial conditions affect the results, we hope to gain a better understanding of the dynamic aspects of opinion formation in specific social contexts and situations.

### 2. invariant magnetic field

Subgroups brought about by items in $F$ $A$ and Influence on $B$ Opinion Polarity:. Individuals under the influence of $F$ may have their opinion polarity strengthened or reversed depending on its value. For example, a positive external influence will increase the likelihood that the individual's opinion will be positive. Conversely, a negative influence will make the opinion negative.

**Speed of Change in Opinion**

The magnitude of $F$ also affects how quickly opinions change. A large external influence can cause an opinion to change rapidly.

**Intergroup influences Groups**

If some individuals in *A* or *B* are subject to an external influence, that influence may spread to other group members. This is caused by interactions based on trust matrices.

## 2. Trust-Distrust Model, Opinion Distribution

Opinion dynamics theory is applied to compute simulations of human behavior in society. In this paper, we introduce distrust into the bounded trust model in order to discuss the time transition and trust between the two. For a fixed agent, $1 \leq i \leq N$, the agent's opinion at time $t$ is $I_i(t)$. As a trust coefficient, we modified the meaning of the coefficient $D_{ij}$ in the bounded trust model. Here, we assumed that $D_{ij} > 0$ if there is trust between them and $D_{ij} < 0$ if there is distrust between them. For the calculations in this paper, $D_{ij}$ was assumed to be constant. Thus, the change in the opinion of agent $i$ can be expressed as follows:

$$\Delta I_i(t) = -\alpha I_i(t) + c_i A(t)\Delta t + \sum_{j=1}^{N} D_{ij} I_j(t) \Delta t \quad (1)$$

$$D_{ij}\phi(I_i, I_j)(I_j(t) - I_i(t)) \quad (2)$$

$$\phi(I_i, I_j) = \frac{1}{1 + \exp(\beta(|I_i - I_j| - b))} \quad (3)$$

Here, in order to cut off the influence from people whose opinions differ significantly(3), we use the following sigmoidal smooth cutoff function, which is a Fermi function system. In other words, the model hypothesizes that people do not pay attention to opinions that are far from their own. We will have a separate discussion on the introduction regarding this Fermi function system in the future. Here, $D_{ij}$ and $D_{ji}$ are assumed to be independent. Usually, $D_{ij}$ is an asymmetric matrix.

**Three Groups(Include Subgroup) Equations**

$$\Phi(u, \beta, \alpha, j, i) = \frac{1}{1 + \exp\left(\beta\left(|u[j] - u[i]| - \alpha\right)\right)}$$
$$= \frac{1}{1 + \exp\left(\beta\left(|u[j] - u[i]|\right) - \beta\alpha\right)} \quad (4)$$

Here, $\Phi$ is a function that calculates the weight influencing the exchange of opinions based on the relative difference between opinions, with the following parameters:

$\beta$: A parameter that adjusts the sensitivity in the exchange of opinions.

$\alpha$: A threshold indicating how different two opinions are.

$$o[i]_{\text{next}} = o[i] + h\left(a + \sum_{j \neq i} d[i,j](o[j] - o[i])\Phi(o, \beta, \alpha, j, i) - d[i,i]o[i] + F[i]\right) \quad (5)$$

Where,

$o[i]$: The current opinion of individual $i$.

$d[i,j]$: An element of the trust matrix indicating the trust or distrust between individuals $i$ and $j$.

$F[i]$: The external influence on individual $i$.

**Parameters**

$n_A, n_B, n_F$:
$$n = n_A + n_B + n_F$$

These represent the number of individuals in groups A, B, and F, respectively.

$o$: An array indicating the initial opinions of each individual.

$$o = (\text{np.random.rand}(n) \times 2 - 1) \times 20$$

$F$: An array indicating the external influence on each individual.

$d$: The trust matrix. Each element indicates the trust or distrust between two individuals.

$\beta, \alpha$: Parameters of the $\Phi$ function.

$z$: The total number of simulation steps.

$h$:
$$h = \frac{t_{\max} - t_{\min}}{z}$$

The time interval of each simulation step.

$a$:
$$a = 0.005 \times \text{np.random.rand}()$$

A small value representing random noise.

**1. Initial Opinions**

$$o_i = (2 \times \text{rand}() - 1)(t)$$

Where:

$o_i$ is the initial opinion of the $i$th individual.

rand(): A function generating a uniform random number between 0 and 1.

## 2. External Influence

$$F_k = \begin{cases} A & \text{for } k < 10 \\ B & \text{for } n_A \leq k < n_A + 10 \\ C & \text{for } n_A + n_B \leq k < n \end{cases}$$

Where:

$F_k$ represents the external influence on the $k$th individual.

## 3. Trust Matrix

Within groups:

$$d_{ij} = \begin{cases} \text{rand}() & \text{if rand}() < \text{trust\_percent} \\ -\text{rand}() & \text{otherwise} \end{cases}$$

For individuals between group F and the other groups:

$$d_{ij} = \begin{cases} \text{rand}() & \text{if rand}() < \text{trust\_val} \\ -\text{rand}() & \text{otherwise} \end{cases}$$

Where:

$d_{ij}$ is an element of matrix $d$ indicating the strength of trust relationship between individual $i$ and individual $j$.

trust_percent: Probability of a positive trust relationship.

trust_val: Trust value towards external influence from group F (varies depending on whether it's for group A, group B, or group F).

## 4. Time Step Size

$$h = \frac{tmax - tmin}{z}$$

Where:

$h$ is the time step size.

$tmin, tmax$: Starting and ending time of the simulation, respectively.

$z$: Total number of time steps.

### (1) Factors that can contribute to reasoning about social phenomena

Initial Opinion (Initial Value)The opinions of individuals within a society are diverse, and this diversity acts as an initial condition for opinion formation and change. This initial opinion is intended to reflect an individual's prior experience and access to information on a particular topic.

External influences indicate factors that are not considered part of society but have a significant impact on opinion formation, such as the influence of external sources of information, opinion leaders, and the media. It is intended to assume the degree of influence of a particular information source or opinion leader.

The trust matrix indicates the strength and direction of interactions between individuals. It can be thought of as an indicator of the cohesiveness or fragmentation of a society. It is also intended to infer the level of social trust as it indicates how likely one is to trust (or distrust) others.

Time step size (h): this parameter indicates the time resolution of the model. It is used to understand the rate at which opinions fluctuate and the speed of social responses.

### (2) Influence of subgroups on $A$ and $B$ caused by items in the invariant field $F$.

The value of the direction of influence determines how members of subgroups $A$ and $B$ are affected. For example, If $F_k$ is positive, then the opinion of subgroup $A$ may be reinforced if the opinion is positive. On the other hand, if subgroup B's opinion is negative, its opinion may be reversed.

Trust influence: The influence of subgroup F depends on how much the members of subgroups $A$ and $B$ trust subgroup $F$. A higher level of trust will result in a stronger influence, while a lower level of trust will result in less influence.

### (3) Social Consequences

If the opinions and beliefs of subgroup $F$ strongly influence subgroups $A$ and $B$, this would potentially indicate that subgroup $F$ has a strong position as an opinion leader and information source. Such information is the case for information campaigns and opinion formation strategies.

In this proposal, by explicitly incorporating interactions among $AB_s$, $AF_s$, and $BF_s$ in the trust matrix, we have added elements that can contribute to the inference of social phenomena as follows.

## Factors that can contribute to the Inference of Social Phenomena

The strength and directional $D_{ij}$ elements of interactions between groups allow us to assess how much members of a particular group $A$ are influenced by the opinions and information of members of group $B$ and outside group $F$, and vice versa. This is used to understand and infer the degree and direction of opinion exchange between different social groups and subgroups.

### (1)Subgroup Influence.

If there is a high level of trust between groups $A$ and $B$, or between groups $A$ and $F$, or between group $B$ and the unchanging position: subgroup $F$, then the opinions and information of that group may have a significant influence on other groups. This is an indicator of how much influence a particular group has in shaping social opinion.

### (2) Propagation and acceptance of information

By analyzing the elements of the trust matrix, it is possible to infer how information propagates among groups and which groups are more likely to accept the information. This provides a basis for evaluating the efficiency and effectiveness of information propagation.

### (3) Possibility of social fragmentation

If the elements of the trust matrix are low between a given group, it indicates that the two groups are likely to distrust each other. We hope to infer the cause of social division and conflict from such information.

### (4) Interaction with outside influences

If subgroup $F$ plays a role in bringing in outside opinions and information, we hope that by clarifying the degree and direction of interaction between that group and other groups ($A$ and $B$), we can infer the extent to which information and opinions from outside influence the formation of opinions in society as a whole.

## 3. Discussion

The result of this calculation is $A_t$ timestep 100: Number of individuals who changed their opinion from $A$ to $B$: 0 Number of individuals who changed their opinion from $A$ to $B$: 0 Number of individuals who changed their opinion from $B$ to $A$: 6

The above is for $A = 100$ and $B = 100$, with an initial confidence value of 80%. The trust relationship between the invariant field subgroup F and $A$, $B$ is

$trust_{FF} = 0.6$, $trust_{FA} = 0.7$, $trust_{FB} = 0.5$ due to the following factors.

### 1. the process of consensus building under conditions of trust and distrust between $A$ and $B$

The trust matrix indicates the level of trust between groups and between individuals within a group. The speed and direction of convergence of opinions depend on the level of trust indicated by this matrix. If the level of trust is high, the group is more likely to be influenced by the opinions of the group, and opinions are more likely to converge. Conversely, if distrust is high, it is inferred that there is an increased tendency to distance oneself from the opinions of that group. The model assumes that Group $A$ and Group $B$ have different initial opinions. If the difference in initial opinions is large, it may take longer to reach agreement.

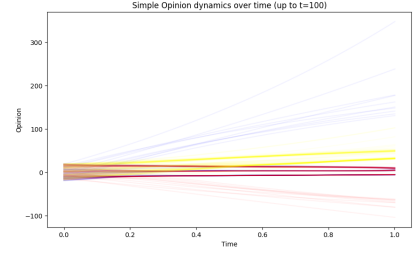

Fig. 1: Simple Opinion Dynamics between the invariant field subgroup F, $t = 100$

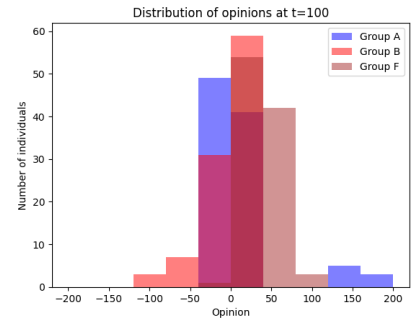

Fig. 2: Distribution of opinion between the invariant field subgroup F, $t = 100$

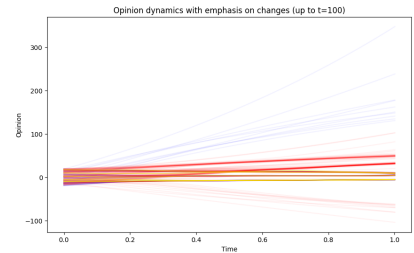

Fig. 3: Emphasis on changes between the invariant field subgroup F, $t = 100$

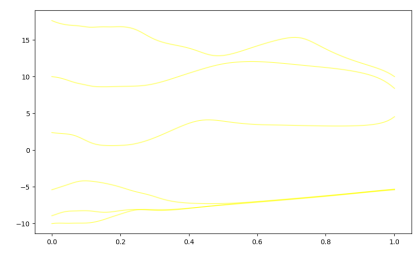

Fig. 4: The trust relationship between the invariant field subgroup F, $t = 100$

## 2. what is the case and opinion formation in society?

Trust between groups may be based on political, economic, and cultural backgrounds in the real world. Examples include trust between communities of different religions and cultures, and trust between business partners. The process of Group *A* and Group *B* agreeing can be interpreted as a process of negotiation, dialogue, and consensus building within a society. Examples could include international negotiations, business transactions, and the exchange of opinions in local communities.

## 3. considerations from $trust_{FF}$, $trust_{FA}$, and $trust_{FB}$

The invariant magnetic field F can be interpreted as an external influence or source of information. It can be considered as a factor that influences the opinions of certain groups or individuals, such as the media, celebrities, or government.

**(a) consensus building process.**

When the value of $trust_{FA}$ or $trust_{FB}$ is high, Group *A* or Group *B* is more likely to accept information or influence from outside sources. This may lead to a faster change and convergence of opinions.

**(b) What is the case and opinion formation in society?**

If the value of $trust_{FA}$ or $trust_{FB}$ is high, it can be interpreted as Group A or Group B having high trust in external information sources (e.g. media or government). Conversely, if these values are low, it can be interpreted as a high level of suspicion or distrust of external information sources. This situation may be based on the political context, the level of trust in the media, or cultural and historical background.

The result of this calculation is $A_t$ timestep the 700: Number of individuals who changed their opinion from *A* to *B*: 0 Number of individuals who changed their opinion from *A* to *B*: 0 Number of individuals who changed their opinion from *B* to *A*: 37

The above is for $A = 100$ and $B = 100$, with an initial confidence value of 80%. The trust relationship between the invariant field subgroup F and *A*, *B* is

**(1) The process of consensus building under conditions of trust and distrust between *A* and *B***

With the trust matrix set up, there is an 80% probability that a trust relationship has been established between groups A and B. Therefore, opinions within the groups are more likely to be influenced by each other. However, the results show that there were no observed changes in opinions from Group *A* to *B*, and 37 changes from Group *B* to *A*. This suggests that a combination of initial opinions, external influences, and other factors make Group B more likely to change its opinion.

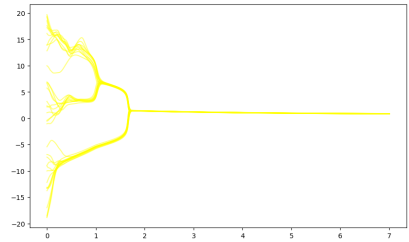

Fig. 5: The trust relationship between the invariant field subgroup F, $t = 100$

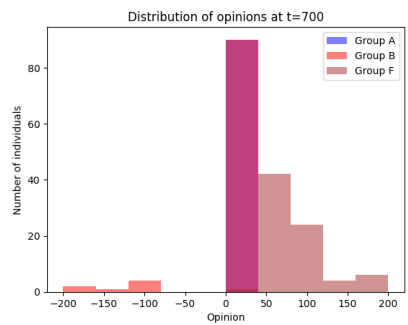

Fig. 6: Distribution of opinion between the invariant field subgroup F, $t = 100$

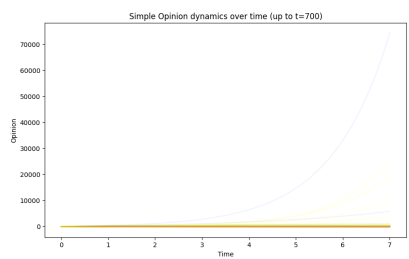

Fig. 7: Emphasis on changes between the invariant field subgroup F, $t = 100$

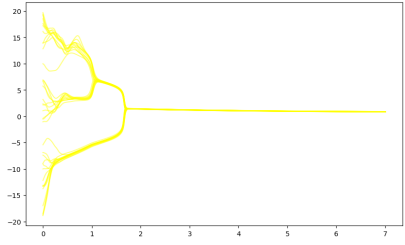

Fig. 8: Emphasis on changes between the invariant field subgroup F, $t = 100$

This model can be interpreted as an indication of the process of opinion formation within a social network or community. For example, it models how people's opinions on a particular topic, such as political positions or product preferences, change depending on their trust relationships and external sources of information. The results show that some communities and groups are more likely to be influenced by outside information and the opinions of other groups. On the other hand, another group is shown to be less susceptible to this influence.

**(2) Considerations from** $trust_{FF} = 0.6$ , $trust_{FA} = 0.7$ , $trust_{FB} = 0.5$

**(a) As a consensus building process**

Group $A$ has 70% trust in external source $F$, while Group $B$ has only 50% trust. This indicates that Group $A$ is more receptive to information from outside sources. The result that people in Group $B$ shift to the views of Group $A$ suggests that outside source F has a stronger influence on Group $A$.

**(b) What is the case and opinion formation in society**

The model shows how external sources of information, for example media and influencers, influence a particular community or group. For example, it suggests that a group that trusts one source of information may change its opinion based on the information provided by that source and influence other groups to do so.

From the above Fig.8, let us consider the divergence of opinions and the process of consensus building between groups $A$ and $B$ as a model of social phenomenon as follows.

**(1) The process of consensus building under conditions of trust and distrust between $A$ and $B$**

Consensus building occurs as a result of interactions within the community and external influences; when groups $A$ and $B$ have different levels of trust, convergence of opinions can be asymmetric. When a high level of trust is set between groups $A$ and $B$, it is easier for them to influence each other's opinions, and their opinions are more likely to come closer together. However, if changes in opinion are observed in only one direction, this may indicate that Group B's opinion is more susceptible to Group $A$, or that Group $A$ is under stronger external influence.

**(2) The case of opinion formation in society**

In society, opinion formation depends heavily on the reliability of information sources, social ties, and individual beliefs and values. Groups with a high level of trust in a particular source of information are more likely to accept information from that source. On the other hand, social ties and trust have been shown to play an important role in opinion convergence. The process by which opinions diverge depends on the balance of social influences, information flows, and communication patterns. The divergence of opinion as illustrated in the model may apply to many real-world situations, such as social debates, political movements, and market trends. The acceptance of an opinion or product within a particular group may spread through that group, but this will depend largely on trust between groups and outside influences.

## 4. Perspects

Next, we need to begin a discussion on $A_{(}t)$ external efficacy and the impact of invariant opinion location $F$, etc. Here, we will first note only the formulas and parameters as an idea note, and then look forward to the future.

Given a set of individuals $i \in \{1, \ldots, n\}$, the opinion of individual $i$ at time $t + 1$, denoted as $o_i(t + 1)$, is updated according to the influence of other individuals and an external force $F$. The update rule is as follows:

$$o_i(t+1) = o_i(t) + h\left(a + \sum_{j \neq i} d_{ij}(o_j - o_i)\Phi(o, \beta, \alpha, j, i) - d_{ii}o_i(t) + F_i\right)$$

where: - $o_i(t)$ is the current opinion of individual $i$. - $h$ is the time step size. - $a$ is a small random factor to model spontaneous opinion changes. - $d_{ij}$ is the trust score between individuals $i$ and $j$. - $\Phi(o, \beta, \alpha, j, i)$ is a sigmoid function modeling the influence of opinion difference. - $F_i$ is the influence of the external force on individual $i$.

The trust matrix $d$, the opinion vector $o$, and the external force vector $F$ are initialized as follows:

$d_{ij}$ = random trust score between -1 and 1,

$o_i$ = initial opinion drawn from a uniform distribution over $[-20, 20]$,

$$F_i = \begin{cases} A & \text{for the first 10 individuals of group A,} \\ B & \text{for the first 10 individuals of group B after A,} \\ C & \text{for the individuals of group F.} \end{cases}$$

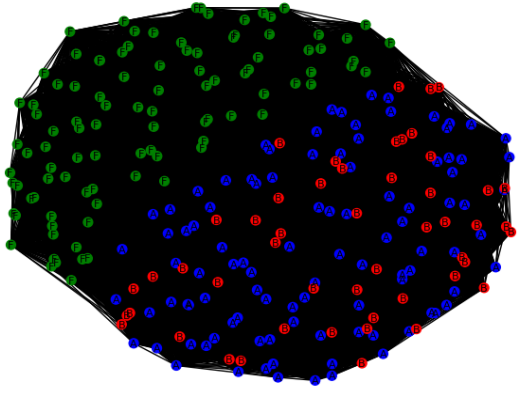

Fig. 9: Network of opinion between the invariant field subgroup F, $t = 800$

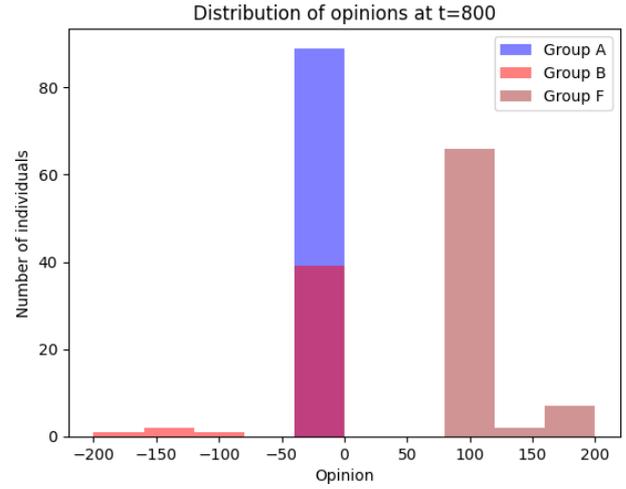

Fig. 10: Distribution of opinion between the invariant field subgroup F, $t = 800$

The trust towards the external force is modeled by adjusting the trust scores within the trust matrix based on the group affiliations and predefined trust levels towards the external force.

**Opinion Dynamics Model with External Influence**

**Description**

Consider a population divided into three groups: A, B, and F. Let $n_A, n_B, n_F$ denote the number of individuals in each group, and $n = n_A + n_B + n_F$ be the total population. Each individual $i$ holds an opinion $o_i(t)$, which evolves over time according to the following update rule:

**Network Analysis**

The network of trust relationships is constructed by connecting individuals with positive trust scores. The structure of this network is crucial for understanding how information and opinions propagate through the population.

**Societal Implications**

The model reflects the complexity of opinion dynamics in a society where individuals are influenced by both their social connections and external forces. It demonstrates the potential for opinions to be swayed or reinforced, leading to phenomena such as polarization or consensus. The differing levels of trust towards external influences among groups $A$, $B$, and $F$ indicate varying degrees of susceptibility to such forces, which can have profound implications for social harmony and the democratic process.

**Pre-Results**

Figure 9, Figure 10 shows, Pre-Results provided simulates the dynamics of opinion formation in a population consisting of three groups: A, B, and F, with an external influence represented by A(t). The external influence is characterized by its impact on the opinions of individuals within each group, with varying levels of trust towards this force. The simulation includes interactions within and between these groups under the influence of an external force, which is less affected by spatial (or network) distances.

Here is an outline to discuss the mathematical framework and social implications based on the Pre-Results provided:

**Trust Matrix and External Influence**

The trust matrix $d$ is tailored to reflect trust levels within and between groups and towards an external force $F$. The force $F$ is modeled to have a uniform effect irrespective of the network distance.

**Network Construction**

The social network is constructed with nodes representing individuals and edges for positive trust relationships ($d_{ij} > 0$). This allows for the analysis of group connectivity and clustering.

**Consensus Building Process**

Introduction of $A_{(t)}$ external influence: In this model, external influence is considered as an integral part of opinion formation, which is particularly evident in the high level of trust in the $F$ group. In the presence of external influences, individual opinions are more likely to accept outside opinions in the process of consensus building through interaction, which may result in greater homogenization of opinions in society.

**Trust and Distrust: Program has a matrix of trust Relationships**

$D_{ij}$, which plays an important role in the opinion formation process. The higher the level of trust, the stronger the influence of others on the opinions of others, which is thought to promote consensus building.

**Opinion Formation in Society**

Conflict between groups: Low levels of trust between *A* and *B* groups, or varying levels of trust within a group with respect to external influences, can accentuate differences in opinion between groups and lead to conflict.

**Information Filter Bubble**

High levels of trust within the same group can create information bubbles or echo chamber effects. This creates a situation where in-group opinions are self-reinforcing to the exclusion of other group opinions.

Concentration of Social Influence: Groups with high levels of trust in outside influences, such as the *F* group, are more likely to have concentrated influence in society. In this situation, a few sources can be expected to have a large impact.

Opinion polarization: When distrust is high, opinions may become more polarized and divided. Different groups may disagree with each other, deepening the division of society as a whole.

## 5. Conclusion

This analysis for capturing the complexity of opinion formation in society, we trying. Through the dynamics of trust and distrust, we will can understand how interactions between groups shape and change opinions. When external influence is strong, trust in a particular source or authority may be concentrated and opinions may be homogenized. On the other hand, increased distrust between groups can lead to increased division and polarization of opinion. Using such a model, this study aims to analyze social phenomena in order to better understand the mechanisms of opinion formation.

## Aknowlegement

The author is grateful for discussion with Prof. Serge Galam. This research is supported by Grant-in-Aid for Scientific Research Project FY 2019-2021, Research Project/Area No. 19K04881, "Construction of a new theory of opinion dynamics that can describe the real picture of society by introducing trust and distrust".